%% file: 4U1705-32_MAP.tex
\documentclass[a4paper,fleqn,usenatbib]{mnras}

\usepackage{newtxtext,newtxmath}
\usepackage[T1]{fontenc}
\usepackage{ae,aecompl}

\usepackage{graphicx}	% Including figure files
\usepackage{amsmath}	% Advanced maths commands
\usepackage{amssymb}	% Extra maths symbols
\usepackage{multicol}        % Multi-column entries in tables
\usepackage{bm}		% Bold maths symbols, including upright Greek
\usepackage{pdflscape}	% Landscape pages

\usepackage{epsfig}
\usepackage{threeparttable} 
\usepackage{epstopdf}
\usepackage[usenames, dvipsnames]{color}

\usepackage{booktabs,array}

\usepackage{color}

\newcommand{\RXS}{1RXS~J170854.4-321857}

\include{definitions}

\title[The ultra-compact candidate \RXS]{On the ultra-compact nature of the neutron star system \RXS: insights from X-ray spectroscopy}

\author[M. Armas Padilla and E. L\'opez-Navas]{
M. Armas Padilla,$^{1,2}$%\thanks{E-mail: m.armaspadilla@iac.es}
and E. L\'opez-Navas,$^{1,2,3}$
\\
$^{1}$Instituto de Astrof\'isica de Canarias (IAC), V\'ia L\'actea s/n, La Laguna 38205, S/C de Tenerife, Spain\\
$^{2}$Departamento de Astrof\'isica, Universidad de La Laguna, La Laguna, E-38205, S/C de Tenerife, Spain\\
$^{3}$Instituto de F\'isica y Astronom'\'ia, Facultad de Ciencias, Universidad de Valpara\'iso, Gran Bretana N 1111, Playa Ancha, Valpara\'iso, Chile
}

\date{Accepted XXX. Received YYY; in original form ZZZ}

\pubyear{2019}

\begin{document}
\label{firstpage}
\pagerange{\pageref{firstpage}--\pageref{lastpage}}
\maketitle

\begin{abstract}
The relatively small family of ultra-compact X-ray binary systems is of great interest for many areas of astrophysics. We report on a detailed X-ray spectral study of the persistent neutron star low mass X-ray binary \RXS. We analysed two \xmm\ observations obtained in late 2004 and early 2005 when, in agreement with previous studies, the system displayed an X-ray luminosity (0.5--10~keV) of $\sim1\times10^{36}\lum$. The spectrum can be described by a Comptonized emission component with $\Gamma\sim$1.9 and a distribution of seed photons with a temperature of $\sim0.23$~keV. A prominent residual feature is present at soft energies, which is reproduced by the absorption model if over-abundances of Ne and Fe are allowed. We discuss how similar observables, that might be attributed to the peculiar (non-solar) composition of the plasma donated by the companion star, are a common feature in confirmed and candidate ultra-compact systems. Although this interpretation is still under debate, we conclude that the detection of these features along with the persistent nature of the source at such low luminosity and the intermediate-long burst that it displayed in the past confirms \RXS as a solid ultra-compact X-ray binary candidate.

\end{abstract}


%%%%%%%%%%%%%%%%%%%%%%%%%%%%%%%%%%%%%%%%%%%%%%%%%%

%%%%%%%%%%%%%%%%% BODY OF PAPER %%%%%%%%%%%%%%%%%%

%%%%%%%%%%Table LOG %%%%%%%%%%%%

\begin{table*}
\centering
\caption{ \xmm\ observations log.}
\begin{threeparttable}
\begin{tabular}{ccccccc}
\hline 
ID			& Date					& Exposure					&EPIC Camera			& Net exposure				& Net count rate \\
   			&(yyyy-mm-dd)			&[ks]						&					&[ks]						&[$\cnts$]\\
\hline
0206990201	&2004-10-01 				&  13.4					 	&MOS1				&11.6						& 1.05$\pm$0.01\\
  			&						&							&MOS2				&12.	1						& 0.98$\pm$0.01\\
  			\\
0206991101	&2005-02-19 				&  12.9						&MOS1				&10.2						&1.5$\pm$0.01 \\
   			&						&							&MOS2				&10.7						&2.1$\pm$0.01\\
    			&						&							&PN					&7.6							&4.3$\pm$0.03 \\
\hline
  
\end{tabular}%
\label{tab:log}
\end{threeparttable}
\end{table*}

%%%%%%%%%%%%%%%%%%%%%%%%%%%%%%%%%%%
%%%%%%%%%%%%%%%%%%%%%%%%%%%%%%%%%%%

\section{Introduction}\label{sec:intr}

The vast majority of the known galactic population of stellar-mass black holes (BHs) and a significant fraction of the neutron stars (NS) are found in low mass X-ray binaries (LMXBs). These stellar systems have sub--solar companion stars that transfer material to the compact object via Roche lobe overflow. Among LMXBs, the population of ultra-compact X-ray binaries (UCXBs), comprised by systems with orbital periods shorter than 80~min, is of particular interest. These short periods imply small Roche lobes, in which only degenerated (hydrogen poor) donor stars can fit. UCXBs are unique laboratories to study accretion processes in hydrogen deficient environments as well as some of the fundamental stages of binary evolution \citep[e.g., common-envelope phase,][]{Nelemans2009,Nelemans2010,Tauris2018}. Last but not least, UCXBs will be primary sources for gravitational waves studies at low-frequencies by the forthcoming LISA mission \citep{Nelemans2018,Tauris2018}.  

Although the predicted number of UCXBs in our galaxy is $\sim10^{6}$ \citep{Nelemans2009}, at present only 14 systems have been confirmed (i.e. with reported orbital periods; see e.g., \citealt{Strohmayer2018}; \citealt{Heinke2013}, and references therein). One of the reasons of this scarcity is that measuring the orbital periods is not always straightforward. Standard techniques include the search for periodic eclipses/modulations for sources seen at high inclination or Doppler-delayed pulses in the case of pulsar acretors. In the optical regime, modulations due to accretion disc super-humps or triggered by irradiation of the companion star also enable to infer orbital solutions.  When direct measurements are not feasible, we need to use indirect evidences to identify UCXB candidates. 

\citet{IntZand2007} proposed a method based on the disc instability model (DIM; see \citealt{Lasota2001} for a review) that identifies systems persistently accreting at very low luminosities as potential UCXBs. Indeed, a fraction of the now confirmed UCXB population were initially candidates proposed by this method. In addition, optical spectroscopic studies have been used to test the ultra-compact nature of several systems. This is done by investigating the absence/presence of emission/absorption features in the spectra, which provide hints on the chemical composition of the accreted material, and thus, on the nature of the companion star \citep{Nelemans2004,Nelemans2006,IntZand2008, Santisteban2017}. Likewise, X-ray spectroscopy has been used to search for UCXB candidates. This is based on the presence of a common feature at $\sim$0.7--1~keV, attributed to an enhanced Ne/O ratio in the plasma donated by the (white dwarf) companion star \citep{Juett2001, Juett2003b}. Despite this technique reported successfully confirmed candidates, it has to be taken with caution as the Ne/O ratio was observed to vary with luminosity in some sources \citep[][see section \ref{sec:Disc}]{Juett2005}

\RXS\ is a NS LMXB that has been detected by several X-ray missions since the early 70's \citep[][and references therein]{IntZand2005}. The source has shown a persistently low unabsorbed flux of a few times $10^{-11}~\flux$ \citep{IntZand2004,IntZand2005}, while only near-infrared upper-limits of $\gtrsim$20 mags (J, K and I bands; \citealt{Revnivtsev2013}) have been reported. An intermediate-long type~I X-ray burst ($\sim$10~min) showed mild photospheric radius expansion, from which -- assuming either a pure hydrogen or helium atmosphere -- a distance of 13$\pm$2~kpc was derived \citep{IntZand2005}. This translates into a persistent luminosity of \lx(0.5--10~keV)$\sim1\times10^{36}~\lum$. This persistently low \lx, together with the aforementioned detection of the 10-min burst, which occur predominantly on UCXBs (albeit not exclusively; see e.g., \citealt{Galloway2017, IntZand2019}) suggested that the system is an UCXB \citep{IntZand2005,IntZand2007}. In this work, we present a detailed  \xmm\  spectral study that provides additional support to the ultra-compact nature of \RXS.

\section{Observations and data reduction}\label{sec:obs}

The \xmm\ observatory \citep{Jansen2001} pointed at \RXS\ on 2004 January 1 and 2005 February 19 for $\sim$12~ksec, respectively. Only the two Metal Oxide Semi-conductor cameras \citep[MOS1 and MOS2][]{Turner2001} of the European Photon Imaging Camera (EPIC) were active during the first observation, while the PN camera \citep{Struder2001} was also available for the second pointing (see Table \ref{tab:log} for the observing log). The detectors were operated in imaging (full-frame window) mode with the medium optical blocking filter during both observations. We extract calibrated events and scientific products using the Science Analysis Software (\textsc{sas}, v.16.0.0). We excluded episodes of background flaring by removing data with high-energy count rates $>$~0.3 and $>$~0.5~$\cnts$ for MOS and PN cameras, respectively. We extracted the source events using a circular region of 75~arcsec radius excising a radius of 15~arcsec (MOS1 and MOS2 for the 2004 observation),  10~arcsec (MOS1 and PN for the 2005 observation) and 5~arcsec (MOS2 of the 2005 observation) in order to mitigate pile-up effects. Background events were extracted using a circular region of 100~arcsec radius placed in a source-free region of the CCDs. We selected events with pixel pattern  below 5 and 13 for PN and MOS detectors, respectively.
We extracted light curves and spectra and generated response matrix files (RMFs) and ancillary response files (ARFs) following the standard analysis threads\footnote{\url{https://www.cosmos.esa.int/web/xmm-newton/sas-threads}}. We rebinned the spectrum in order to include a minimum of 25 counts in every spectral channel and avoiding to oversample the full width at half-maximum of the energy resolution by a factor larger than 3. We used \textsc{xspec} (v.12.9.1; \citealt{Arnaud1996}) to analyse the \xmm\ spectra.

Finally, we also inspected and extracted the data from the Reflection Grating Spectrometers \citep[RGS;][]{Herder2001}, but we did not include them in the  in our analysis due to their low statistics.

\section{Analysis and results}\label{sec:res}

We investigated the light curves in order to search for features revealing the orbital period, such as eclipses or absorption dips \citep{White1995}. We created a 100s-bin light-curve using the PN data, from which we derive a upper limit on the mean aperiodic varibility of <5\%, a constraint that it is limited by the quality of our data.  A visual inspection to the (2$\times$3~h) light-curves do not show any clear high inclination feature. In order to test this further, we performed a flux vs hardness diagram (not shown; see e.g. fig 5 in \citealt{Kuulkers2013}), which does not reveal any clear trend, as it would be expected for the case of variable absorption (e.g. dips). Thus, we conclude that either \RXS\ has an orbital inclination $i\lesssim65\degr$ \citep{White1985, Cantrell2010} or that the orbital period is longer than 3.5~h. The latter scenario, given the X-ray characteristics of the source, seems unlikely.

%
%%%%%%%%%Figure XMM Spectra %%%%%%%%%%%%
\begin{figure*}$\phantom{!t}$
\centering
\begin{tabular}{cc}
%\begin{center}
\includegraphics[keepaspectratio,width=\columnwidth, trim=0.0cm 0.0cm 0.0cm 0.0cm]{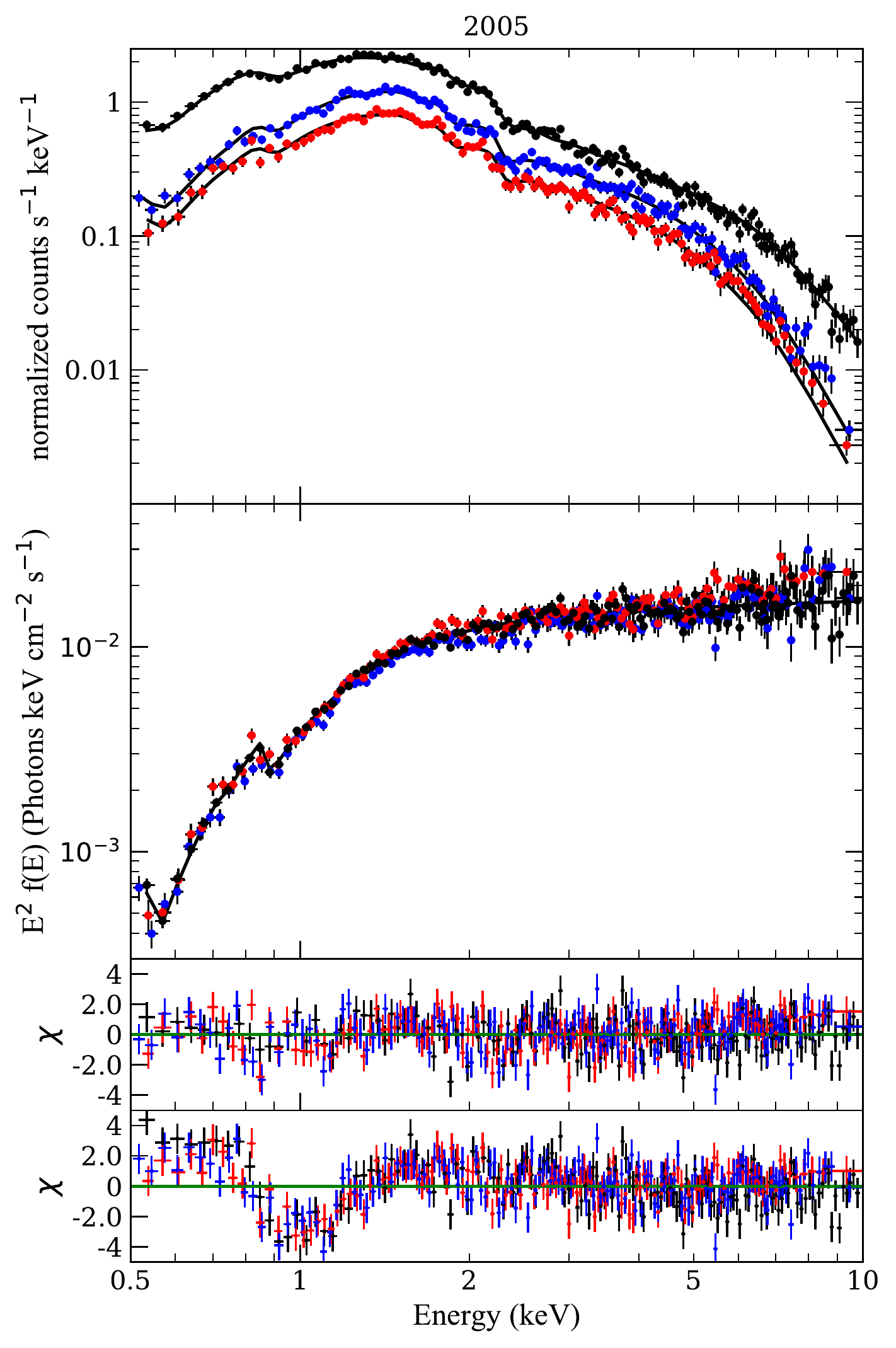}
&
\includegraphics[keepaspectratio,width=\columnwidth, trim=0.0cm 0.0cm 0.0cm 0.0cm]{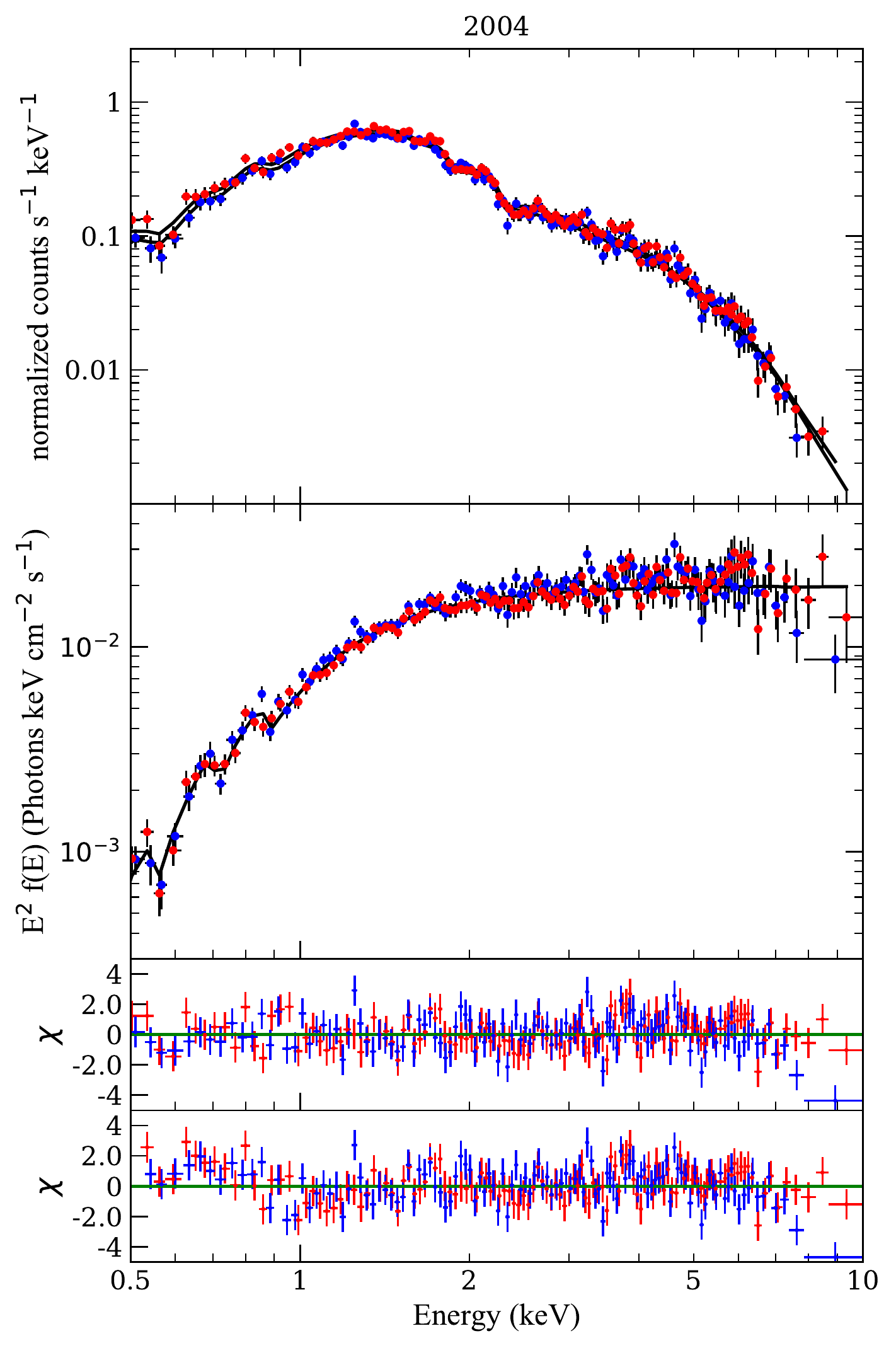}\\
\end{tabular}
\caption{PN (black), MOS1 (red) and MOS2 (blue) spectra taken in 2005 (left) and 2004 (right). From top to bottom, the first panel shows the folded spectra while the second panel depicts the unfolded data. The fit using TBNEW+NTHCOMP is shown as a black solid line. The corresponding fit residuals in units of $\sigma$ are shown in the third panel. The bottom panels shows the fit residuals in units of $\sigma$ when using TBABS+NTHCOMP (i.e., Solar abundances).}

\label{fig:spectra}
%\end{center}
\end{figure*}

%%%%%%%%%%%%%%%%%%%%%
%%%%%%%%%%%%%%%%%%%%%

We started our spectral study by analysing the 2005--observation, which includes the three EPIC cameras and hence has better statistics. We simultaneously fitted the 0.5--10~keV PN, MOS1 and MOS2 spectra with the parameters tied between the tree detectors. In order to account for cross-calibration uncertainties between the cameras, we included a constant factor (CONSTANT) in our models. We fix it to 1 for the PN spectrum and leave it free to vary for the MOS spectra. In our first attempt, we used a single thermally Comptonized continuum model (NTHCOMP in \textsc{xspec}; \citealt{Zdziarski1996, Zycki1999}) modified by photoelectric absorption which we modelled using the Tuebingen-Boulder Interstellar Medium absorption model (TBABS in \textsc{xspec}) with cross-sections of \citet{Verner1996} and abundances of \citet{Wilms2000}. The corona electron temperature parameter ($kT_{\rm e}$) typically adopts values above 10~keV in the low-hard state of LMXBs, which is beyond our 0.5--10~keV spectral coverage. Therefore, we fixed $kT_{\rm e}$ to 25~keV, which is the typical value displayed by NS systems at this low-luminous states \citep{Burke2017}. We note that variations of this value do not have an impact on the results (within errors). This simple absorbed Comptonization model was not able to reproduce our data. The fit residuals exhibited a strong feature around $\sim$1~keV (see left plot on Fig. \ref{fig:spectra}, bottom panel) with \qhis\ of 661 for 416 degrees of freedom (dof) (p-value\footnote{The probability value (p-value) represents the probability that the deviations between the data and the model are due to chance alone. In general, a model can be rejected when the p-value is smaller than 0.05.} of $2\times10^{-13}$). We added a thermal component (DISKBB or BBODYRAD) assuming that the soft residuals might be produced by emission from the accretion disc or NS surface/boundary layer, as commonly observed in NSs accreting a low luminosities \citep[e.g.][]{ArmasPadilla2017, ArmasPadilla2018}. However, the extra thermal component does not account for the soft residuals and an evident structure remains below 2~keV. 

Soft excesses below $\sim$1~keV  have been observed  in  spectra of highly obscured X-ray binaries obtained in EPIC-PN Timing mode (see XMM-SOC-CAL-TN-0083 \footnote{\url{ http://xmm.vilspa.esa.es/docs/documents/CAL-TN-0083.pdf}}). In some cases this feature has been attributed to residual uncertainties in the redistribution calibration \citep{Hiemstra2011}. However, it is unlikely that this is the cause of our residuals since (i) our spectra are taken in Imaging mode, (ii) the source is only mildly absorbed, and (iii) the structure is present in all three EPIC detectors. On the other hand, similar features have been observed in several (candidate) UCXBs \citep[e.g.][]{Juett2001,Juett2003b,Farinelli2003, IntZand2008}. These are proposed to result from over-abundances in the absorbing material, generally enhanced O, Ne and Fe, which is probably intrinsic to the sources. Following the same procedure reported in the literature, we replaced the TBABS model by the absorption model TBNEW, which allows the abundances to vary \citep{IntZand2008,Madej2014,VandenEijnden2018}. We started again with a simple absorbed Comptonization model (i.e., TBNEW*NTHCOMP).
We kept fixed all the TBnew parameters at their default values, except the equivalent hydrogen column density (\Nh). We tried several different fits by allowing both individual elements and combination of elements (e.g., C, O, Fe, Ne, etc) to vary. We found that soft residuals were present to some extend unless the abundances of Ne and Fe were allowed to vary freely. Indeed, this combination provided the best fit, which significantly improved previous attempts ($\Delta$\qhis=341 for 3 dof with respect to the same model with Solar abundances). Although we are aware that the fit is still poor from a purely statistical point of view (p-value is still $<$0.05), we can not decrease the \qhired\ by adding extra continuum components (e.g., DISKBB, BBODYRAD) to our model. Moreover, there is not any other feature in our residuals clearly suggesting that a component is missed. Therefore, we concluded that the impossibility of improvement might be caused by remaining uncertainties in the cross-calibration between the different detectors (\citealt{Kirsch2004}; XMM-SOC-CAL-TN-0052). We obtained an \Nh\ of (3.4$\pm$0.4)$\times 10^{21}$\nh (consistent with the reported value by \citealt{IntZand2005} using \chan\ data) and over-abundances of Ne  ($A_{\rm Ne}$=5.4$\pm$0.4) and Fe ($A_{\rm Fe}$=2.5$\pm$0.7). The Comptonization asymptotic power-law photon index ($\Gamma$) is $1.91\pm0.02$ and the temperature of the up-scattered seed photons ($kT_{\rm seed}$) assuming black body geometry is 0.23$\pm$0.06~keV (consistent values are obtained assuming a disc geometry). The inferred 0.5--10~keV unabsorbed flux of (6.3$\pm$0.2)$\times10^{-11}~\flux$ is implies a luminosity of \lx$\sim1.7\times10^{36}\lum$ assuming a distance of 13~kpc \citep{IntZand2005}. We report best-fit results in Table \ref{tab:res} (see also left panel in Fig.\ref{fig:spectra}). Uncertainties are given at 90 per cent confidence level.

For the 2004 observation, that only includes MOS1 and MOS2 data, we fixed the constant (cross-calibration) factor to 1 for the former and leave it free to vary for MOS2. We fixed \Nh\ to the value obtained in the best fit of our higher signal--to--noise 2005 observation (\Nh=3.4$\times 10^{21}$\nh). As for the 2005 data, a simple absorbed Comptonization model was not able to reproduce the data. This could not be solved by adding soft thermal components (\qhis$\cong$319 for 239~dof; p-value=$4\times10^{-4}$). Although it is less evident than for the 2005 observation, some residuals below 1~keV are also present (see bottom--right panel in Fig.\ref{fig:spectra}). Thus, we repeated the fit using TBNEW*NTHCOMP with free Ne and Fe abundances. This produced an acceptable fit, with \qhis= 275 for 243~dof (p-value$\cong$0.07). The fit and corresponding residuals are shown in Fig.\ref{fig:spectra} (right plot). The resulting spectral values are consistent with those obtained from the 2005 observation (Table \ref{tab:res}), albeit the system was slightly brighter in 2004. %(8.5$\pm$0.2)$\times10^{-11}~\flux$.

%%%%%%%%%%Table RESULTS %%%%%%%%%%%%
\begin{table}
\centering
\caption{Fitting results for the 2004 and 2005 data using the TBNEW*NTCHOMP model. Uncertainties are expressed at 90 per cent confidence level.}
\begin{threeparttable}
\begin{tabular}{ l c c }
\hline
  Component 							&	2004 			& 	2005	 \\		 			
\hline
$C_{\rm PN}$							& -- 						&   1 (fix)		\\
$C_{\rm MOS1}$						& 1 (fix)					&   1.04$\pm$0.02 		\\
$C_{\rm MOS2}$						& 1.05$\pm$0.02	&   0.94$\pm$0.01		\\
\Nh\ ($\times 10^{22}$ \nh)			& 	0.34 (fix) 		&  0.34$\pm$0.4			 		\\
$A_{\rm Ne}$							& 	4.1$\pm$1		&  5.4$\pm$0.4 		 \\
$A_{\rm Fe}$							& 	3.5$\pm$1		&  2.5$\pm$0.7  		 \\
$\Gamma$				 				& 	1.98$\pm$0.04   & 1.91$\pm$0.02 		   \\
$kT_{\rm e}$ (keV)					& 	25 (fix)	 		& 25 (fix)	 	\\
$kT_{\rm seed}$ (keV)				& 	0.21$\pm$0.02	& 0.23$\pm$0.03  	\\
$N_{\rm nthcomp}$ ($\times 10^{-2}$) & 	1.6$\pm$0.1		& 1.06$\pm$0.1 	  	 			  	\\
\qhis\ (dof) 						& 	283(249) 			& 522 (414)					  \\
 &&\\
\multicolumn{1}{r}{(0.5--10 keV)} 	&&\\
\Fx\ ($\times10^{-11} \flux$)		&	8.5$\pm$0.2		& 6.3$\pm$0.2			 	\\
\lx$^{\rm a}$ ($\times10^{36}\lum$)	&	1.72	$\pm$0.04		& 1.27$\pm$0.04				 \\

\hline
\end{tabular}
\begin{tablenotes}
\item[a]{X-ray luminosity assuming a distance of 13~kpc.}

\end{tablenotes}
\label{tab:res}
\end{threeparttable}
\end{table}
%%%%%%%%%%%%%%%%%%%%%%%%%%%%%%%%%%%
%%%%%%%%%%%%%%%%%%%%%%%%%%%%%%%%%%%

%%%%%%%%%%Table UCXBs with 0.7 structure %%%%%%%%%%%%
\begin{table*}
\centering
\caption{List of confirmed and candidate UCXBs with peculiar soft X-ray structures.}
\begin{threeparttable}
\begin{tabular}{ l c c c c  c c c  }
\hline
Source 				&Period 		& P/T	 &  \multicolumn{2}{c}{Spectral data}	& Interm-long& \Nh &   \lx\ (Energy band)   \\ %& References
                    & (minutes)  &    &   X-ray   				 	& Optical	&burst          		&$\times10^{21}$\nh				&  $\lum$ (keV)       \\	
\hline

2S 0918-549  		& 17.4 		& P	&			Ne/O			  & He?	&  	 IB$^{a}$		& 3.0	&$\sim3.5\times10^{35}$ (2--10) $^{b}$			\\ %& Juett2003 \\		

4U 1543-624 			&18.2		& P	&			Ne/O			     & C/O	&	 --		&3.5		&$\sim2-4\times10^{36}$ (2--10) $^{b}$    	\\ %juett2003,Farinelli2003\\

4U 1850-087         & 20.6		& P	&			Ne/O			   	&?	&    IB$^{a}$			& 3.9	&$\sim1\times10^{36}$ (0.5--10) $^{c,d}$ \\%	& Juett2001,2005 \\

IGR J17062-6143	   & 37.97$^{e}$	& P	&  			O$^{f}$        		& lack \ha $^{g}$	&    IB$^{a}$      &   2.4$^{h}$   & $\sim2.9\times10^{35}$ (0.3--79)  $^{h}$\\% &  Jakob2018)                     \\

4U 1626-67		   &42 			& P	&		    Ne/O, C, O, Ne  		   			& C/O	&  	 --      &1.4      &       $\sim1\times10^{36}$ (0.5--10)  $^{i}$                             \\ %& Schulz2001    \\
\hline
\multicolumn{8}{c}{Candidates} \\
\hline
4U 0614+091		  &  51?        & P      &       Ne/O, O                     & C/O	&    IB$^{j}$       & 3.0      &    $\sim1\times10^{36}$ (0.5--10) $^{c}$              \\ %&   \\

1A 1246-588  &   ?				& P	&   			Ne$^{k}$			   			& Featureless$^{k}$ &	 IB$^{a}$	&2.5$^{k}$ &   $\sim6\times10^{35}$ (0.3--10) $^{k}$ \\%& \\\citetalt{IntZand2008}

\textbf{\RXS}  &   ?				& P  &   		Ne, Fe				 			& ?	&	 IB$^{l}$	& 3.4	&   $\sim1\times10^{36}$ (0.5--10)\\	%& \\

\hline
\end{tabular}
\begin{tablenotes}
\item{ Based on data presented in \citet{Paradijs1994}, \citet{Nelemans2006}, \citet{VanHaaften2012}, \citet{Heinke2013}.}
\item{References: $^{a}$\citet{IntZand2019}, $^{b}$\citet{Juett2003b}, $^{c}$\citet{Juett2001}, $^{d}$\citet{Juett2005}, $^{e}$\citet{Strohmayer2018}, $^{f}$\citet{VandenEijnden2018}, $^{g}$\citet{Santisteban2017},$^{h}$\citet{Degenaar2017}, $^{i}$\citet{Schulz2001}, $^{j}$\citet{Kuulkers2010}, $^{k}$\citet{IntZand2008}, $^{l}$\citet{IntZand2005},     } 

\end{tablenotes}
\label{tab:UCXB}
\end{threeparttable}
\end{table*}
%%%%%%%%%%%%%%%%%%%%%%%%%%%%%%%%%%%
%%%%%%%%%%%%%%%%%%%%%%%%%%%%%%%%%%%

\section{Discussion}\label{sec:Disc}

\RXS\ is a NS LMXB that has been detected  without exception at \lx$\sim$0.01\ledd~by several missions over the past decades. Within the framework of the disc instability model \citep{Lasota2001,Coriat2012} this persistent nature at such low luminosity implies a short orbital period, possibly within the UCXB regime \citep{IntZand2005}. To investigate this further, we have carried out a detailed analysis of two archival \xmm\ observations. The 0.5--10~keV X-ray luminosity during these observations was $\sim1\times10^{36}\lum$, which is consistent with previous measurements  reinforcing the persistent nature of the system \citep[e.g.,][]{Forman1978,Markert1979, Revnivtsev2004,IntZand2005}. Our spectra are well described by a thermally Comptonized model with $\Gamma\cong1.9$ and a distribution of seed photons characterised by $kT_{\rm seed}\cong0.2$~keV arising from either the accretion disc or NS surface/boundary layer (possibly from both regions; see \citealt{ArmasPadilla2018}). We do not detect direct emission from the accretion disc or NS surface as no softer thermal component was required in our analysis. NS systems accreting at luminosities $<10^{35}\lum$ usually show a soft thermal component, generally attributed to low-level accretion on to the neutron star surface \citep{Zampieri1995,ArmasPadilla2013c,Bahramian2014}. This typically contributes at the $\sim$30--50 per cent level \citep[e.g.,][]{Degenaar2013, ArmasPadilla2013b, Arnason2015, ArmasPadilla2018}. However, at luminosities above $\sim 10^{35}\lum$ this contribution becomes lower or, in some cases, totally vanishes \citep{ArmasPadilla2013b, Allen2015, Wijnands2015}. The non-detection of the thermal component together with the relatively soft photons index ($\Gamma \sim 1.9$) concurs with the latter scenario.

Interestingly, the spectra showed a broad residual structure around $\sim$1~keV, which we were able to account for by allowing relative abundances to vary in the absorption model. We found  \Nh=(3.4$\pm$0.4)$\times 10^{21}$\nh\ with enhanced Ne and Fe abundances. This \Nh\ is consistent with the value obtained by \citet{IntZand2005}, albeit their abundances are in agreement with the standard interstellar value. The reason for this difference might be related to the significantly worse statistics of their spectra (with only 58 dof). We obtain Ne and Fe relative abundances of $A_{\rm Ne}$=5.4$\pm$0.4 and $A_{\rm Fe}$=2.5$\pm$0.7, while a maximum value of $\sim$1.4 have been measured for the interstellar medium towards LMXBs \citep{Pinto2013}. Hence, the additional absorption required to fit both the 2004 and 2005 data is unlikely to have an interstellar origin, being probably intrinsic to the system. Although these over-abundances are rather high, similar values have been reported for some other UCXBs \citep[e.g,][]{IntZand2008, Madej2011}. Nevertheless, we note that due to the relatively low spectral resolution of our data and the systematics likely involved, to derive the actual abundances and species producing  the soft X-ray feature presented here is beyond the scope of the paper. 

A number of confirmed (and candidate) UCXBs have exhibited similar phenomenology to that reported here, which, in most cases, has been interpreted as due to over-abundances in the absorber. This generally involves excesses in the K-edges of neutral O and Ne and the L-edge of neutral Fe,  which origin have been suggested to be intrinsic to the binaries. \citet{Juett2001} first proposed the ultra-compact nature of several systems based on the detection of enhanced Ne/O ratios using X-ray spectra \citep[see also,][]{Juett2003b}. This feature had been originally found in the confirmed UCXBs 4U~1626-67 and 4U~1850-087. In light of  these findings, it was suggested that systems showing this peculiarity harbour C--O or O--Ne--Mg white dwarfs as companions. Also, a white dwarf donor was proposed for IGR~J17062-6143, the last confirmed UCXB to date \citep{Strohmayer2018}. It showed a residual structure at $\sim$1~keV, that was suggested to arise from an over-abundance of O -- likely related with circumbinary material -- while Ne edge abundances were consistent with the interstellar medium \citep{VandenEijnden2018, Degenaar2017}. The presence of strong emission lines of highly ionised species of Ne and O in the high resolution X-ray spectroscopy of 4U~1626-67 also suggested a C--O rich white dwarf donor \citep{Schulz2001,Krauss2007}. Optical spectroscopic studies of some of these sources support the above conclusions. The detection of C and O emission lines evidences the presence of metal--rich material in the accretion disc, consistent with C--O white dwarf donors.  \citep{Nelemans2004,Werner2006,Nelemans2006}. Additionally, the lack of strong H and He emission lines, typically the strongest features in the optical spectra of LMXBs \citep{Charles2006, MataSanchez2018}, supports the ultra-compact nature of some of the systems that have shown the peculiar X-ray features  \citep[e.g.;][]{Nelemans2006,IntZand2008, Santisteban2017}. In this context, it is important to emphasise that the Ne/O ratio was questioned as UCXB marker due to the hydrogen and helium lines found in the optical spectrum of the NS system 4U~1556-60 \citep{Nelemans2006, Nelemans2010}. However, 4U~1556-60 has not shown any of the X-ray features classically associated with uncommon abundances \citep{Farinelli2003}, and therefore, it was not (and has never been as far as we know) an UCXB candidate.

High resolution X-ray spectra of some UCXBs have shown, in addition to over-abundances in the absorber, the  presence of a broad \ion{O}{vii}~Ly$_{\alpha}$ emission line \citep[e.g.][]{Madej2010, Schulz2010, Madej2011}. This, together with the Fe~K$_{\alpha}$ line, is predicted to be the most prominent fluorescent line of the reflection spectrum if X-rays are reprocessed in a oxygen-rich accretion disc. Indeed, reflection models have been modified to account for the non-solar composition of the accretion disc in order to reproduce the reflection spectra of UCXBs \citep{Koliopanos2013,Madej2014}. Moreover, \citet{Madej2014} found that better results were obtained if the ionization structure of the illuminated disc was also taken in account. Likewise, it was suggested that ionization effects might be responsible for the epoch-to-epoch variations of the Ne/O ratio found in some systems \citep{Juett2003b,Juett2005, IntZand2008}. In this scenario, changes in the continuum spectral properties would modify the ionization state of the different (absorber related) species in the accretion disc. Thus, the measured Ne/O ratio would not reflect the donor composition. However, this does not explain why the $\sim$1~keV structure is so common in UCXBs as compared with the whole LMXB population. In Table \ref{tab:UCXB} we list candidate and confirmed UCXBs that, to the best of our knowledge, have shown the soft X-ray features compatible with peculiar Ne and O abundances. Interestingly, these systems have a persistent nature, \Nh $<5\times10^{21}$\nh and were accreting at \lx$\lesssim$0.01\ledd\ at the time they showed the soft feature. 

Finally, numerous UCXBs have displayed intermediate-long type~I X-ray burst, albeit these are not exclusively observed in UCXBs \citep[e.g.][]{Falanga2008}. These thermonuclear explosions are thought to be associated with deep He ignitions from material accreted at very low rates ($\lesssim0.01$~\ledd), suggesting He-rich donors (\citealt{IntZand2005b,Cumming2006,Falanga2008,Kuulkers2010,Galloway2017,IntZand2019}). As a matter of fact, \citet{IntZand2005b} proposed that the anomalous Ne/O ratios could result from a reduction of the O abundance due to CNO processes in helium white dwarfs. This scenario might apply to \RXS. In this regard, the detection of a $\sim$10~min X-ray burst, the persistently low accretion rate ($\sim$0.01~\ledd) and the apparent abundances anomalies that we have reported here agree with the above picture. Higher X-ray spectral resolution data at high signal-to-noise are required in order to further investigate the nature of \RXS\ in particular, and the physical processes behind the abundance-anomalies related features commonly observed in UCXBs.

\section*{Acknowledgements}
MAP acknowledge support by the Spanish MINECO under grant AYA2017-83216-P. MAP's research is funded under the Juan de la Cierva Fellowship Programme (IJCI-2016-30867). ELN acknowledge the IAC Summer Fellowship, during which part of the research leading to this results were carried out. \xmm\ is an ESA science mission with instruments and contributions directly funded by ESA Member States and NASA. 
%We acknowledge the use of public data from the \swift\ data archive. 

%%%%%%%%%%%%%%%%%%%%%%%%%%%%%%%%%%%%%%%%%%%%%%%%%%


\bibliographystyle{mnras}
\bibliography{4U1705-32_MAP.bbl} % if your bibtex file is called example.bib

%%%%%%%%%%%%%%%%%%%%%%%%%%%%%%%%%%%%%%%%%%%%%%%%%%

%%%%%%%%%%%%%%%%% APPENDICES %%%%%%%%%%%%%%%%%%%%%

%\appendix

%%%%%%%%%%%%%%%%%%%%%%%%%%%%%%%%%%%%%%%%%%%%%%%%%%

% Don't change these lines
\bsp	% typesetting comment
\label{lastpage}
\end{document}

%% file: definitions.tex
\newcommand{\ha}{H$\alpha$}

\newcommand{\Fx}{$F_\mathrm{X}$}
\newcommand{\lx}{$L_\mathrm{X}$}
\newcommand{\ledd}{$L_\mathrm{Edd}$}
\newcommand{\Nh}{$N_{\rm H}$}
\newcommand{\qhired}{$\chi^{2}_{\nu}$}
\newcommand{\qhis}{$\chi^{2}$}

\newcommand{\lum}{\mathrm{erg~s}^{-1}}
\newcommand{\flux}{\mathrm{erg~cm}^{-2}~\mathrm{s}^{-1}}
\newcommand{\cnts}{\mathrm{counts~s}^{-1}}
\newcommand{\nh}{$\mathrm{cm}^{-2}$}

\newcommand{\chan}{\textit{Chandra}}

\newcommand{\xmm}{\textit{XMM-Newton}}

%% file: 4U1705-32_MAP.bbl
\begin{thebibliography}{}
\makeatletter
\relax
\def\mn@urlcharsother{\let\do\@makeother \do\$\do\&\do\#\do\^\do\_\do\%\do\~}
\def\mn@doi{\begingroup\mn@urlcharsother \@ifnextchar [ {\mn@doi@}
  {\mn@doi@[]}}
\def\mn@doi@[#1]#2{\def\@tempa{#1}\ifx\@tempa\@empty \href
  {http://dx.doi.org/#2} {doi:#2}\else \href {http://dx.doi.org/#2} {#1}\fi
  \endgroup}
\def\mn@eprint#1#2{\mn@eprint@#1:#2::\@nil}
\def\mn@eprint@arXiv#1{\href {http://arxiv.org/abs/#1} {{\tt arXiv:#1}}}
\def\mn@eprint@dblp#1{\href {http://dblp.uni-trier.de/rec/bibtex/#1.xml}
  {dblp:#1}}
\def\mn@eprint@#1:#2:#3:#4\@nil{\def\@tempa {#1}\def\@tempb {#2}\def\@tempc
  {#3}\ifx \@tempc \@empty \let \@tempc \@tempb \let \@tempb \@tempa \fi \ifx
  \@tempb \@empty \def\@tempb {arXiv}\fi \@ifundefined
  {mn@eprint@\@tempb}{\@tempb:\@tempc}{\expandafter \expandafter \csname
  mn@eprint@\@tempb\endcsname \expandafter{\@tempc}}}

\bibitem[\protect\citeauthoryear{Allen, Linares, Homan  \& Chakrabarty}{Allen
  et~al.}{2015}]{Allen2015}
Allen J.~L.,  Linares M.,  Homan J.,   Chakrabarty D.,  2015, \mn@doi [ApJ]
  {10.1088/0004-637X/801/1/10}, 801, 10

\bibitem[\protect\citeauthoryear{{Armas Padilla}, Degenaar  \& Wijnands}{{Armas
  Padilla} et~al.}{2013a}]{ArmasPadilla2013b}
{Armas Padilla} M.,  Degenaar N.,   Wijnands R.,  2013a, \mn@doi [MNRAS]
  {10.1093/mnras/stt1114}, 434, 1586

\bibitem[\protect\citeauthoryear{{Armas Padilla}, Wijnands  \& Degenaar}{{Armas
  Padilla} et~al.}{2013b}]{ArmasPadilla2013c}
{Armas Padilla} M.,  Wijnands R.,   Degenaar N.,  2013b, \mn@doi [MNRAS Lett]
  {10.1093/mnrasl/slt119}, 436, L89

\bibitem[\protect\citeauthoryear{{Armas Padilla}, Ueda, Hori, Shidatsu  \&
  Mu{\~{n}}oz-Darias}{{Armas Padilla} et~al.}{2017}]{ArmasPadilla2017}
{Armas Padilla} M.,  Ueda Y.,  Hori T.,  Shidatsu M.,   Mu{\~{n}}oz-Darias T.,
  2017, \mn@doi [MNRAS] {10.1093/mnras/stx020}, 309, 290

\bibitem[\protect\citeauthoryear{{Armas Padilla}, Ponti, {De Marco},
  Mu{\~{n}}oz-Darias  \& Haberl}{{Armas Padilla}
  et~al.}{2018}]{ArmasPadilla2018}
{Armas Padilla} M.,  Ponti G.,  {De Marco} B.,  Mu{\~{n}}oz-Darias T.,   Haberl
  F.,  2018, \mn@doi [MNRAS] {10.1093/mnras/stx2538}, 473, 3789

\bibitem[\protect\citeauthoryear{Arnason, Sivakoff, Heinke, Cohn  \&
  Lugger}{Arnason et~al.}{2015}]{Arnason2015}
Arnason R.~M.,  Sivakoff G.~R.,  Heinke C.~O.,  Cohn H.~N.,   Lugger P.~M.,
  2015, \mn@doi [ApJ] {10.1088/0004-637X/807/1/52}, 807, 52

\bibitem[\protect\citeauthoryear{Arnaud}{Arnaud}{1996}]{Arnaud1996}
Arnaud K.,  1996, in Jacoby G.,  Barnes J.,  eds,  Astronomical Society of the
  Pacific Conference Series Vol. 101, Astronomical Data Analysis Software and
  Systems V. p.~17

\bibitem[\protect\citeauthoryear{Bahramian et~al.,}{Bahramian
  et~al.}{2014}]{Bahramian2014}
Bahramian A.,  et~al., 2014, \mn@doi [ApJ] {10.1088/0004-637X/780/2/127}, 780

\bibitem[\protect\citeauthoryear{Burke, Gilfanov  \& Sunyaev}{Burke
  et~al.}{2017}]{Burke2017}
Burke M.~J.,  Gilfanov M.,   Sunyaev R.,  2017, \mn@doi [MNRAS]
  {10.1093/mnras/stw2514}, 466, 194

\bibitem[\protect\citeauthoryear{Cantrell et~al.,}{Cantrell
  et~al.}{2010}]{Cantrell2010}
Cantrell A.~G.,  et~al., 2010, \mn@doi [ApJ] {10.1088/0004-637X/710/2/1127},
  710, 1127

\bibitem[\protect\citeauthoryear{Charles \& Coe}{Charles \&
  Coe}{2006}]{Charles2006}
Charles P.~A.,  Coe M.~J.,  2006, in Lewin W.,  van~der Klis M.,  eds, ,
  Vol.~39, In: Compact stellar X-ray sources. Edited by Walter Lewin {\&}
  Michiel van der Klis..
Cambridge University Press, p.~690 (\mn@eprint {arXiv} {0308020}),
  \mn@doi{10.1177/01461079070370020502}, \url
  {http://adsabs.harvard.edu/abs/2006csxs.book..215C
  http://arxiv.org/abs/astro-ph/0308020}

\bibitem[\protect\citeauthoryear{Coriat, Fender  \& Dubus}{Coriat
  et~al.}{2012}]{Coriat2012}
Coriat M.,  Fender R.~P.,   Dubus G.,  2012, \mn@doi [MNRAS]
  {10.1111/j.1365-2966.2012.21339.x}, 424, 1991

\bibitem[\protect\citeauthoryear{Cumming, Macbeth, Zand  \& Page}{Cumming
  et~al.}{2006}]{Cumming2006}
Cumming A.,  Macbeth J.,  Zand J. J. M. i.~t.,   Page D.,  2006, \mn@doi [ApJ]
  {10.1086/504698}, 646, 429

\bibitem[\protect\citeauthoryear{Degenaar, Wijnands  \& Miller}{Degenaar
  et~al.}{2013}]{Degenaar2013}
Degenaar N.,  Wijnands R.,   Miller J.~M.,  2013, \mn@doi [ApJ]
  {10.1088/2041-8205/767/2/L31}, 767, L31

\bibitem[\protect\citeauthoryear{Degenaar, Pinto, Miller, Wijnands, Altamirano,
  Paerels, Fabian  \& Chakrabarty}{Degenaar et~al.}{2017}]{Degenaar2017}
Degenaar N.,  Pinto C.,  Miller J.~M.,  Wijnands R.,  Altamirano D.,  Paerels
  F.,  Fabian A.~C.,   Chakrabarty D.,  2017, \mn@doi [MNRAS]
  {10.1093/mnras/stw2355}, 464, 398

\bibitem[\protect\citeauthoryear{Falanga, Chenevez, Cumming, Kuulkers, Trap  \&
  Goldwurm}{Falanga et~al.}{2008}]{Falanga2008}
Falanga M.,  Chenevez J.,  Cumming A.,  Kuulkers E.,  Trap G.,   Goldwurm A.,
  2008, \mn@doi [A{\&}A] {10.1051/0004-6361:20078982}, 484, 43

\bibitem[\protect\citeauthoryear{Farinelli et~al.,}{Farinelli
  et~al.}{2003}]{Farinelli2003}
Farinelli R.,  et~al., 2003, \mn@doi [A{\&}A] {10.1051/0004-6361:20030165},
  402, 1021

\bibitem[\protect\citeauthoryear{Forman, Jones, Cominsky, Julien, Murray,
  Peters, Tananbaum  \& Giacconi}{Forman et~al.}{1978}]{Forman1978}
Forman W.,  Jones C.,  Cominsky L.,  Julien P.,  Murray S.,  Peters G.,
  Tananbaum H.,   Giacconi R.,  1978, \mn@doi [ApJS] {10.1086/190561}, 38, 357

\bibitem[\protect\citeauthoryear{Galloway \& Keek}{Galloway \&
  Keek}{2017}]{Galloway2017}
Galloway D.~K.,  Keek L.,  2017, in Belloni T.,  Mendez M.,   Zhang C.,  eds, ,
  Timing Neutron Stars: Pulsations, Oscillations and Explosions.
ASSL, Springer (\mn@eprint {arXiv} {1712.06227}), \url
  {http://arxiv.org/abs/1712.06227}

\bibitem[\protect\citeauthoryear{Heinke, Ivanova, Engel, Pavlovskii, Sivakoff,
  Cartwright  \& Gladstone}{Heinke et~al.}{2013}]{Heinke2013}
Heinke C.~O.,  Ivanova N.,  Engel M.~C.,  Pavlovskii K.,  Sivakoff G.~R.,
  Cartwright T.~F.,   Gladstone J.~C.,  2013, \mn@doi [ApJ]
  {10.1088/0004-637X/768/2/184}, 768, 184

\bibitem[\protect\citeauthoryear{{Hern{\'{a}}ndez Santisteban}
  et~al.,}{{Hern{\'{a}}ndez Santisteban} et~al.}{2017}]{Santisteban2017}
{Hern{\'{a}}ndez Santisteban} J.~V.,  et~al., 2017, MNRAS, 000, 1

\bibitem[\protect\citeauthoryear{Hiemstra, M{\'{e}}ndez, Done, Trigo,
  Altamirano  \& Casella}{Hiemstra et~al.}{2011}]{Hiemstra2011}
Hiemstra B.,  M{\'{e}}ndez M.,  Done C.,  Trigo M.~D.,  Altamirano D.,
  Casella P.,  2011, \mn@doi [MNRAS] {10.1111/j.1365-2966.2010.17661.x}, 411,
  137

\bibitem[\protect\citeauthoryear{Jansen et~al.,}{Jansen
  et~al.}{2001}]{Jansen2001}
Jansen F.,  et~al., 2001, \mn@doi [A{\&}A] {10.1051/0004-6361:20000036}, 365,
  L1

\bibitem[\protect\citeauthoryear{Juett \& Chakrabarty}{Juett \&
  Chakrabarty}{2003}]{Juett2003b}
Juett A. M.~M.,  Chakrabarty D.,  2003, \mn@doi [ApJ] {10.1086/379188}, 599,
  498

\bibitem[\protect\citeauthoryear{Juett \& Chakrabarty}{Juett \&
  Chakrabarty}{2005}]{Juett2005}
Juett A.~M.,  Chakrabarty D.,  2005, \mn@doi [ApJ] {10.1086/430633}, 627, 926

\bibitem[\protect\citeauthoryear{Juett, Psaltis  \& Chakrabarty}{Juett
  et~al.}{2001}]{Juett2001}
Juett A.~M.,  Psaltis D.,   Chakrabarty D.,  2001, ApJ, 560, 59

\bibitem[\protect\citeauthoryear{Kirsch et~al.,}{Kirsch
  et~al.}{2004}]{Kirsch2004}
Kirsch M.,  et~al., 2004, in Hasinger G.,  Turner M.,  eds,  Society of
  Photo-Optical Instrumentation Engineers (SPIE) Conference Series Vol. 5488,
  Society of Photo-Optical Instrumentation Engineers (SPIE) Conference Series.
  pp 103--114, \mn@doi{10.1117/12.549276}

\bibitem[\protect\citeauthoryear{Koliopanos, Gilfanov  \& Bildsten}{Koliopanos
  et~al.}{2013}]{Koliopanos2013}
Koliopanos F.,  Gilfanov M.,   Bildsten L.,  2013, \mn@doi [MNRAS]
  {10.1093/mnras/stt542}, 432, 1264

\bibitem[\protect\citeauthoryear{Krauss, Schulz, Chakrabarty, Juett  \&
  Cottam}{Krauss et~al.}{2007}]{Krauss2007}
Krauss M.~I.,  Schulz N.~S.,  Chakrabarty D.,  Juett A.~M.,   Cottam J.,  2007,
  \mn@doi [ApJ] {10.1086/513592}, 660, 605

\bibitem[\protect\citeauthoryear{Kuulkers et~al.,}{Kuulkers
  et~al.}{2010}]{Kuulkers2010}
Kuulkers E.,  et~al., 2010, \mn@doi [A{\&}A] {10.1051/0004-6361/200913210},
  514, A65

\bibitem[\protect\citeauthoryear{Kuulkers et~al.,}{Kuulkers
  et~al.}{2013}]{Kuulkers2013}
Kuulkers E.,  et~al., 2013, \mn@doi [A{\&}A] {10.1051/0004-6361/201219447},
  552, A32

\bibitem[\protect\citeauthoryear{Lasota}{Lasota}{2001}]{Lasota2001}
Lasota J.-P.,  2001, \mn@doi [MNRAS] {10.1016/S1387-6473(01)00112-9}, 45, 449

\bibitem[\protect\citeauthoryear{Madej \& Jonker}{Madej \&
  Jonker}{2011}]{Madej2011}
Madej O.~K.,  Jonker P.~G.,  2011, \mn@doi [Mon. Not. R. Astron. Soc]
  {10.1111/j.1745-3933.2010.00989.x}, 412, 11

\bibitem[\protect\citeauthoryear{Madej, Jonker, Fabian, Pinto, Verbunt  \& {De
  Plaa}}{Madej et~al.}{2010}]{Madej2010}
Madej O.~K.,  Jonker P.~G.,  Fabian A.~C.,  Pinto C.,  Verbunt F.,   {De Plaa}
  J.,  2010, \mn@doi [MNRAS Lett] {10.1111/j.1745-3933.2010.00892.x}, 407, L11

\bibitem[\protect\citeauthoryear{Madej, Garc{\'{i}}a, Jonker, Parker, Ross,
  Fabian  \& Chenevez}{Madej et~al.}{2014}]{Madej2014}
Madej O.~K.,  Garc{\'{i}}a J.,  Jonker P.~G.,  Parker M.~L.,  Ross R.,  Fabian
  A.~C.,   Chenevez J.,  2014, \mn@doi [MNRAS] {10.1093/mnras/stu884}, 442,
  1157

\bibitem[\protect\citeauthoryear{Markert et~al.,}{Markert
  et~al.}{1979}]{Markert1979}
Markert T.~H.,  et~al., 1979, \mn@doi [ApJS] {10.1086/190587}, 39, 573

\bibitem[\protect\citeauthoryear{{Mata S{\'{a}}nchez} et~al.,}{{Mata
  S{\'{a}}nchez} et~al.}{2018}]{MataSanchez2018}
{Mata S{\'{a}}nchez} D.,  et~al., 2018, \mn@doi [MNRAS]
  {10.1093/mnras/sty2402}, 481, 2646

\bibitem[\protect\citeauthoryear{Nelemans}{Nelemans}{2018}]{Nelemans2018}
Nelemans G.,  2018, arXiv:1807.01060

\bibitem[\protect\citeauthoryear{Nelemans \& Jonker}{Nelemans \&
  Jonker}{2010}]{Nelemans2010}
Nelemans G.,  Jonker P.~G.,  2010, \mn@doi [New Astronomy Reviews]
  {10.1016/j.newar.2010.09.021}, 54, 87

\bibitem[\protect\citeauthoryear{Nelemans, Jonker, Marsh  \& {Van Der
  Klis}}{Nelemans et~al.}{2004}]{Nelemans2004}
Nelemans G.,  Jonker P.~G.,  Marsh T.~R.,   {Van Der Klis} M.,  2004, \mn@doi
  [MNRAS] {10.1111/j.1365-2966.2004.07486.x}, 348, L7

\bibitem[\protect\citeauthoryear{Nelemans, Jonker  \& Steeghs}{Nelemans
  et~al.}{2006}]{Nelemans2006}
Nelemans G.,  Jonker P.~G.,   Steeghs D.,  2006, \mn@doi [MNRAS]
  {10.1111/j.1365-2966.2006.10496.x}, 370, 255

\bibitem[\protect\citeauthoryear{Nelemans et~al.,}{Nelemans
  et~al.}{2009}]{Nelemans2009}
Nelemans G.,  et~al., 2009, Decadal white paper, 2010, 1

\bibitem[\protect\citeauthoryear{Pinto, Kaastra, Costantini  \& de Vries}{Pinto
  et~al.}{2013}]{Pinto2013}
Pinto C.,  Kaastra J.~S.,  Costantini E.,   de Vries C.,  2013, \mn@doi
  [A{\&}A] {10.1051/0004-6361/201220481}, 551, A25

\bibitem[\protect\citeauthoryear{Revnivtsev et~al.,}{Revnivtsev
  et~al.}{2004}]{Revnivtsev2004}
Revnivtsev M.,  et~al., 2004, \mn@doi [Astronomy Letters] {10.1134/1.1764884},
  30, 382

\bibitem[\protect\citeauthoryear{Revnivtsev, Kniazev, Karasev, Berdnikov  \&
  Barway}{Revnivtsev et~al.}{2013}]{Revnivtsev2013}
Revnivtsev M.~G.,  Kniazev A.,  Karasev D.~I.,  Berdnikov L.,   Barway S.,
  2013, \mn@doi [Astronomy Letters] {10.1134/s1063773713080082}, 39, 523

\bibitem[\protect\citeauthoryear{Schulz, Chakrabarty, Marshall, Canizares, Lee
  \& Houck}{Schulz et~al.}{2001}]{Schulz2001}
Schulz N.~S.,  Chakrabarty D.,  Marshall H.~L.,  Canizares C.~R.,  Lee J.~C.,
  Houck J.,  2001, \mn@doi [ApJ] {10.1086/323988}, 563, 941

\bibitem[\protect\citeauthoryear{Schulz, Nowak, Chakrabarty  \&
  Canizares}{Schulz et~al.}{2010}]{Schulz2010}
Schulz N.~S.,  Nowak M.~A.,  Chakrabarty D.,   Canizares C.~R.,  2010, \mn@doi
  [ApJ] {10.1088/0004-637X/725/2/2417}, 725, 2417

\bibitem[\protect\citeauthoryear{Strohmayer et~al.,}{Strohmayer
  et~al.}{2018}]{Strohmayer2018}
Strohmayer T.~E.,  et~al., 2018, \mn@doi [The Astrophysical Journal Letters]
  {10.3847/2041-8213/aabf44}, 858, L13

\bibitem[\protect\citeauthoryear{Str{\"{u}}der et~al.,}{Str{\"{u}}der
  et~al.}{2001}]{Struder2001}
Str{\"{u}}der L.,  et~al., 2001, \mn@doi [A{\&}A] {10.1051/0004-6361:20000066},
  365, L18

\bibitem[\protect\citeauthoryear{Tauris}{Tauris}{2018}]{Tauris2018}
Tauris T.~M.,  2018.  (\mn@eprint {arXiv} {arXiv:1809.03504v1}), \url
  {https://xxx.lanl.gov/pdf/1809.03504v1}

\bibitem[\protect\citeauthoryear{Turner et~al.,}{Turner
  et~al.}{2001}]{Turner2001}
Turner M.,  et~al., 2001, \mn@doi [A{\&}A] {10.1051/0004-6361:20000087}, 365,
  L27

\bibitem[\protect\citeauthoryear{Verner, Ferland, Korista  \& Yakovlev}{Verner
  et~al.}{1996}]{Verner1996}
Verner D.,  Ferland G.,  Korista K.,   Yakovlev D.,  1996, \mn@doi [ApJ]
  {10.1086/177435}, 465, 487

\bibitem[\protect\citeauthoryear{Werner, Nagel, Rauch, Hammer  \&
  Dreizler}{Werner et~al.}{2006}]{Werner2006}
Werner K.,  Nagel T.,  Rauch T.,  Hammer N.~J.,   Dreizler S.,  2006, \mn@doi
  [A{\&}A] {10.1051/0004-6361:20053768}, 450, 725

\bibitem[\protect\citeauthoryear{White \& Mason}{White \&
  Mason}{1985}]{White1985}
White N.~E.,  Mason K.~O.,  1985, \mn@doi [Space Sci. Rev.]
  {10.1007/BF00212883}, 40, 167

\bibitem[\protect\citeauthoryear{White, Nagase  \& Parmar}{White
  et~al.}{1995}]{White1995}
White N.~E.,  Nagase F.,   Parmar A.~N.,  1995, in Lewin W. H.~G.,  van
  Paradijs J.,   van~den Heuvel E. P.~J.,  eds, , X-ray binaries.
Cambridge University Press, pp 1--57, \url
  {http://cdsads.u-strasbg.fr/abs/1995xrbi.nasa....1W}

\bibitem[\protect\citeauthoryear{Wijnands, Degenaar, {Armas Padilla},
  Altamirano, Cavecchi, Linares, Bahramian  \& Heinke}{Wijnands
  et~al.}{2015}]{Wijnands2015}
Wijnands R.,  Degenaar N.,  {Armas Padilla} M.,  Altamirano D.,  Cavecchi Y.,
  Linares M.,  Bahramian A.,   Heinke C.~O.,  2015, \mn@doi [MNRAS]
  {10.1093/mnras/stv1974}, 454, 1371

\bibitem[\protect\citeauthoryear{Wilms, Allen  \& McCray}{Wilms
  et~al.}{2000}]{Wilms2000}
Wilms J.,  Allen A.,   McCray R.,  2000, \mn@doi [ApJ] {10.1086/317016}, 542,
  914

\bibitem[\protect\citeauthoryear{Zampieri, Turolla, Zane  \& Treves}{Zampieri
  et~al.}{1995}]{Zampieri1995}
Zampieri L.,  Turolla R.,  Zane S.,   Treves A.,  1995, \mn@doi [ApJ]
  {10.1086/175223}, 439, 849

\bibitem[\protect\citeauthoryear{Zdziarski, Johnson  \& Magdziarz}{Zdziarski
  et~al.}{1996}]{Zdziarski1996}
Zdziarski A.,  Johnson W.,   Magdziarz P.,  1996, MNRAS, 283, 193

\bibitem[\protect\citeauthoryear{Zycki, Done  \& Smith}{Zycki
  et~al.}{1999}]{Zycki1999}
Zycki P.~T.,  Done C.,   Smith D.~A.,  1999, \mn@doi [MNRAS]
  {10.1046/j.1365-8711.1999.02885.x}, 309, 561

\bibitem[\protect\citeauthoryear{den Herder, Brinkman, Kahn,
  Branduardi-Raymont, Thomsen  \& Aarts}{den Herder et~al.}{2001}]{Herder2001}
den Herder J.,  Brinkman A.,  Kahn S.,  Branduardi-Raymont G.,  Thomsen K.,
  Aarts H.,  2001, \mn@doi [A{\&}A] {10.1051/0004-6361:20000058}, 365, L7

\bibitem[\protect\citeauthoryear{{in 't Zand} et~al.,}{{in 't Zand}
  et~al.}{2004}]{IntZand2004}
{in 't Zand} J.,  et~al., 2004, \mn@doi [Nuclear Physics B - Proceedings
  Supplements] {10.1016/j.nuclphysbps.2004.04.083}, 132, 486

\bibitem[\protect\citeauthoryear{{in 't Zand}, Cornelisse  \& M{\'{e}}ndez}{{in
  't Zand} et~al.}{2005a}]{IntZand2005}
{in 't Zand} J. J.~M.,  Cornelisse R.,   M{\'{e}}ndez M.,  2005a, \mn@doi
  [A{\&}A] {10.1051/0004-6361:20052955}, 440, 287

\bibitem[\protect\citeauthoryear{{in 't Zand}, Cumming, van~der Sluys, Verbunt
  \& Pols}{{in 't Zand} et~al.}{2005b}]{IntZand2005b}
{in 't Zand} J. J.~M.,  Cumming A.,  van~der Sluys M.~V.,  Verbunt F.,   Pols
  O.~R.,  2005b, \mn@doi [A{\&}A] {10.1051/0004-6361:20053002}, 441, 675

\bibitem[\protect\citeauthoryear{{in 't Zand}, Jonker  \& Markwardt}{{in 't
  Zand} et~al.}{2007}]{IntZand2007}
{in 't Zand} J. J.~M.,  Jonker P.~G.,   Markwardt C.~B.,  2007, \mn@doi
  [A{\&}A] {10.1051/0004-6361:20066678}, 465, 953

\bibitem[\protect\citeauthoryear{{in 't Zand} et~al.,}{{in 't Zand}
  et~al.}{2008}]{IntZand2008}
{in 't Zand} J. J.~M.,  et~al., 2008, \mn@doi [A{\&}A]
  {10.1051/0004-6361:200809361}, 485, 183

\bibitem[\protect\citeauthoryear{{in 't Zand}, Kries, Palmer  \& Degenaar}{{in
  't Zand} et~al.}{2019}]{IntZand2019}
{in 't Zand} J. J.~M.,  Kries M. J.~W.,  Palmer D.~M.,   Degenaar N.,  2019,
  \mn@doi [A{\&}A] {10.1051/0004-6361/201834270}, 621, A53

\bibitem[\protect\citeauthoryear{van Haaften, Voss  \& Nelemans}{van Haaften
  et~al.}{2012}]{VanHaaften2012}
van Haaften L.~M.,  Voss R.,   Nelemans G.,  2012, \mn@doi [A{\&}A]
  {10.1051/0004-6361/201118067}, 543, A121

\bibitem[\protect\citeauthoryear{van Paradijs \& McClintock}{van Paradijs \&
  McClintock}{1994}]{Paradijs1994}
van Paradijs J.,  McClintock J.,  1994, A{\&}A, 290, 133

\bibitem[\protect\citeauthoryear{van~den Eijnden et~al.,}{van~den Eijnden
  et~al.}{2018}]{VandenEijnden2018}
van~den Eijnden J.,  et~al., 2018, \mn@doi [MNRAS] {10.1093/mnras/stx3224},
  475, 2027

\makeatother
\end{thebibliography}
